\documentclass[preprint,showpacs,amsmath]{revtex4}

\usepackage{times}
\usepackage{hyperref}

\usepackage{graphicx}
\usepackage{dcolumn}
\usepackage{bm}

\begin{document}
\title{
Numerical precision radiative corrections to the Dalitz plot of light and heavy quark unpolarized baryon semileptonic
decays. The cases $\Xi^0 \to \Sigma^+ e\overline{\nu}$ and $\Lambda_c^+ \to \Lambda e^+ \nu$
}

\author{
Rub\'en Flores-Mendieta
}
\affiliation{
Instituto de F{\'\i}sica, Universidad Aut\'onoma de San Luis Potos{\'\i}, \'Alvaro Obreg\'on 64, Zona Centro, San
Luis Potos{\'\i}, S.L.P.\ 78000, Mexico
}

\author{
J.\ J.\ Torres
}
\affiliation{
Escuela Superior de C\'omputo del IPN, Apartado Postal 75-702, M\'exico, D.F.\ 07738, Mexico
}

\author{
M.\ Neri, A.\ Mart{\'\i}nez
}
\affiliation{ Escuela Superior de F\'{\i}sica y Matem\'aticas del IPN, Apartado Postal 75-702, M\'exico, D.F.\ 07738,
Mexico
}

\author{
A.\ Garc{\'\i}a
}
\affiliation{
Departamento de F{\'\i}sica, Centro de Investigaci\'on y de Estudios Avanzados del IPN, Apartado Postal 14-740,
M\'exico, D.F.\ 07000, Mexico
}

\date{\today}

\begin{abstract}
We propose and discuss a numerical use of our previous precision results for the radiative corrections to unpolarized
spin one-half baryon semileptonic decays, which is not compromised to fixing the form factors at prescribed values.
We present various crosschecks and comparisons with other results available in the literature of such analytical
radiative corrections. Our analysis, however, is general and applies to all charge assignments to the baryons allowed
by heavy quarks. The procedure is exemplified with the processes $\Xi^0 \to \Sigma^+ e\overline{\nu}$ and
$\Lambda_c^+ \to \Lambda e^+ \nu$.
\end{abstract}


\maketitle

\section{Introduction}

The analysis of high statistics experiments of spin one-half baryon semileptonic decays requires the inclusion of
precision radiative corrections (RC). Despite the important progress achieved in the understanding of the fundamental
interactions with the standard model \cite{part}, the calculation of RC is still committed to model dependence and so
are the experimental analyses which use these calculations. Fortunately, up to order $(\alpha/\pi)(q/M_1)$ one can show
\cite{low,sirlin,tun91,tun93} that all the model dependence of RC can be absorbed into the already present form factors
and that the remaining part is model independent. The measurement of such effective form factors is then model
independent. The results obtained for the RC can be extended to cover all the six possible charge assignments of the
baryons allowed when heavy quarks are involved \cite{rfm02}.

Although one can assert that the RC can be calculated reliably and once and for all to a high degree of precision,
there still remains the question of their practical use in an actual Monte Carlo simulation. It is this question
which we shall address in this paper. In practice it is convenient to introduce RC into the simulation in a numerical
form. However, our previous results are analytical \cite{tun91,tun93}. Other results in numerical form found in the
literature \cite{toth} are committed to fixing the form factors at given values. This last may be not too bad an
approximation in hyperon semileptonic decays, but it is unacceptable when heavy quarks are involved because the
Cabibbo theory \cite{part} no longer provides a reliable guidance. It is therefore desirable to be able to produce
numerical RC which allow the form factors to be varied freely in the simulation. The goal of this paper is to propose
a method to do this last and also to illustrate it.

In Sec.~II we briefly review our previous results for the RC of a neutral decaying baryon (NDB) and of a negatively
charged decaying baryon (CDB). We shall not reproduce the long detailed analytical results. They can be found in
Refs.~\cite{tun91,tun93}. We shall only reproduce in Appendix A a convenient rearrangement of several terms which
cannot be found in those references. Before proceeding towards our goal, in Sec.~III we shall present three tables
with numerical crosschecks of the results of Refs.~\cite{tun91} and \cite{tun93} and a comparison with the results of
Ref.~\cite{toth}. The reasons for doing this are to help the user gain confidence in the so long analytical
results and to provide him/her with numbers which serve as reference. In Sec.~IV we discuss how to produce arrays of
numerical RC which keep the form factors uncompromising. As an illustration and also to provide the user with
reference numbers we shall produce two tables with numerical arrays. Section V is reserved for a summary and
conclusions.

\section{Radiative corrections to the Dalitz plot}

First we recall our notation and conventions and next we briefly review the RC to the Dalitz plot of unpolarized
baryon semileptonic decays.

The uncorrected transition amplitude for the decay $A\to Bl\overline{\nu}_l$ is
\begin{eqnarray}
{\mathsf M}_0 = \frac{G_V}{\sqrt 2} [\overline{u}_B(p_2) W_\mu(p_1,p_2) u_A(p_1)] [\overline{u}_l(l) O_\mu v_\nu
(p_\nu)], \label{MO}
\end{eqnarray}
where
\begin{eqnarray}
W_\mu (p_1,p_2) & = & f_1(q^2) \gamma_\mu + f_2(q^2) \sigma_{\mu \nu} \frac{q_\nu}{M_1} + f_3(q^2) \frac{q_\mu}{M_1}
\nonumber \\
&  & \mbox{} + \left[g_1(q^2) \gamma_\mu + g_2(q^2) \sigma_{\mu \nu} \frac{q_\nu}{M_1} + g_3(q^2) \frac{q_\mu}{M_1}
\right] \gamma_5. \label{Wmu}
\end{eqnarray}
Here $O_\mu=\gamma_\mu(1+\gamma_5)$, $q=p_1-p_2$ is the four-momentum transfer, $f_i$ and $g_i$ are the Dirac vector
and axial-vector form factors, respectively, and our conventions for the $\gamma $-matrices are those of
Refs.~\cite{tun91,tun93}. All the form factors are assumed to be real in this work. In addition to the virtual RC to
$A\to Bl\overline{\nu}_l$ one must include the bremsstrahlung RC from the decay $A\to Bl\overline{\nu}_l \gamma$.
The four-momenta and masses of the particles involved in these processes will be denoted by $p_1=(E_1,\mathbf{p}_1)$
and $M_1$, $p_2=(E_2,\mathbf{p}_2)$ and $M_2$, $l=(E,\mathbf{l})$ and $m$, $p_\nu=(E_\nu^0,\mathbf{p}_\nu)$ and
$m_\nu=0$, and $k=(\omega,\mathbf{k})$ and $m_\gamma=0$, respectively. The direction of a vector $\mathbf{p}$ will be
denoted by the unit vector ${\hat {\mathbf p}}$. $p_2$, $l$, and $p_\nu$ will also denote the magnitudes of the
corresponding three-momenta when we specialize our calculation to the center-of-mass frame of $A$. No confusion is
expected because in this situation our expressions will not be manifestly covariant. When a real photon $\gamma$ is
emitted, the neutrino energy becomes $E_\nu=E_\nu^0-\omega$.

We shall assume that, even if there is no provision to detect real photons, these latter can be experimentally
discriminated by energy and momentum conservation. Therefore, our expressions will be limited to the RC to the three
body region of the Dalitz plot of the uncorrected decay \cite{tun91,tun93}.

The complete expression for the Dalitz plot with virtual and bremsstrahlung RC containing terms up to order
$(\alpha/\pi)(q/M_1)$ can be compactly expressed as
\begin{equation}
d\Gamma_i = d\Omega \left[A_0^\prime + \frac{\alpha}{\pi} {\Theta_{iI}} \right]. \label{dgi}
\end{equation}
The phase space factor is $d\Omega =(G_V^2/2)\{dE_2dEd\Omega_2d\varphi_l/(2\pi)^5\}2M_1$. The index $i=C,N$
covers the CDB $A^-\to B^0l^-\overline{\nu}_l$ and the NDB $A^0\to B^+l^-\overline{\nu}_l$ cases, respectively. The
full RC are contained in the functions $\Theta_{iI}$. They are given either in terms of triple integrations over the
photon variables, to be performed numerically, or in an analytical form, where all such integrations have been
performed explicitly. We shall not reproduce all the long expressions contained in $\Theta_{iI}$. Instead we shall
give the referencing necessary to find their analytical detailed expressions in our previous work. We shall make one
exception: some expressions can be reduced substantially. For the sake of completeness, these expressions are given
explicitly in Appendix A.

Still in a compact form, the RC for the CDB case are
\begin{equation}
\Theta_{CI} = B_1^\prime(\Phi_C + I_{C0}) + B_{C1}^{\prime\prime} \Phi_C^\prime + C_A^\prime, \label{TCI}
\end{equation}
$B_1^\prime$, $\Phi_C$, $\Phi_C^\prime$, and $B_{C1}^{\prime\prime}$ are found in Eqs.~(11), (6), (7) and (12) of
Ref.~\cite{tun91}, $I_{C0}$ is found in Eq.~(52) of Ref.~\cite{rfm97}. $\Theta_{CI}$ is infrared finite, the infrared
divergence cutoff of the virtual RC contained in $\Phi_C$ is cancelled away by its counterpart in the $I_{C0}$ of the
bremsstrahlung RC. Here it is important to remark that $I_{C0}$ is equivalent to the $I_0(\alpha)$ of Eq.~(26) plus
the first and second summands of Eq.~(27) in Ref.~\cite{tun91}. Thus the last summand in Eq.~(27) together with
Eqs.~(28)-(30) of Ref.~\cite{tun91} give the triple integration form of $C_A^\prime$. Its analytical form is found in
Eqs.~(43)-(45) of this reference. After some algebraic rearrangements, it can be written as
\begin{equation}
C_A^\prime = H_0^\prime\theta_0 + \sum_{i=2}^{16} H_i^\prime\theta_i, \label{CAp}
\end{equation}
the explicit forms of these $H_i^\prime$ are given in Appendix A. $H_1^\prime$ does not appear in $C_A^\prime$. It is
identified with the $B_1^\prime$ of the first term in Eq.~(\ref{TCI}), i.e., $H_1^\prime=B_1^\prime$.

The compact form of the RC in the NDB case is
\begin{equation}
\Theta_{NI} = B_1^\prime (\Phi_N + I_{N0}) + B_{N1}^{\prime\prime} \Phi_N^\prime + C_A^\prime + C_{NA}^\prime,
\label{TNI}
\end{equation}
$B_1^\prime$ and $C_A^\prime$ in this equation are the same of Eq.~(\ref{TCI}). $\Phi_N$ and $\Phi_N^\prime$
are found in Eqs.~(7) and (8) of Ref.~\cite{tun93}, respectively, once the identifications $\Phi_N=2\textrm{Re}
\phi $ and $\Phi_N^\prime=2m\textrm{Re}\phi^\prime$ are made. Also in this reference the $I_{N0}$ is given in
Eq.~(40) and $B_{N1}^{\prime \prime}$ must be identified with $A_1^\prime$ of Eq.~(15). As in the charged case, the
infrared cutoff contained in $\Phi_N$ and $I_{N0}$ has been cancelled away. The triple integration form of
$C_{NA}^\prime$ is found in Eqs.~(38), (39), (44), and (45) of Ref.~\cite{tun93}. Its analytical form is found in
Eqs.~(48)-(50) of this reference. However, it can still be reduced into
\begin{equation}
C_{NA}^\prime = D_1\rho_{N1} + D_2 \rho_{N2}, \label{CNAp}
\end{equation}
where $D_1={f_1^\prime}^2+3{g_1^\prime}^2$, $D_2={f_1^\prime}^2-{g_1^\prime}^2$, and the functions $\rho_{N1}$ and
$\rho_{N2}$ are given in Appendix A.

Expressions (\ref{TCI}) and (\ref{TNI}) are model independent, finite in the infrared and ultraviolet limits, and
are valid to order $(\alpha/\pi)(q/M_1) $. All the model dependence amounts to six constants which have been absorbed
in the form factors of Eq.~(\ref{Wmu}). This fact is indicated by putting primes on such form factors wherever they
appear. Thus, $A_0^\prime$ of Eq.~(\ref{dgi}) has the same form as in the uncorrected case. Explicitly, it is found
in Eq.~(10) of Ref.~\cite{tun91}.

\section{Crosschecks}

Before proceeding towards our goal it is convenient to help the reader gain confidence in our previous analytical
results, as well as to compare them with other results available in the literature.

Although the triple integration form of the RC in Eq.~(\ref{dgi}) is not practical for its use in a Monte Carlo
simulation, it is very useful to crosscheck the so long and tedious analytical form of such corrections. For this
latter purpose, we make numerical comparisons of both forms at fixed values of $E$ and $E_2$ over the Dalitz plot.
The form factors must also be fixed at predetermined values. All these crosschecks were satisfactory and it is not
necessary to display all the details here. Accordingly, we shall present a minimum of illustrative cases. We shall
discuss the processes $\Sigma^-\to ne \overline{\nu}$ and $\Xi^0 \to \Sigma^+ e\overline{\nu}$ as examples
of CDB and NDB decays.

The results are displayed in Tables I and II. The form factors, at zero-momentum transfer, have been given the
arbitrary values $f_1 = 1.0$, $f_2 = -0.97$, $f_3 = 0.789$, $g_1 = -0.34$, $g_2 = -1.567$, and $g_3 = 0.766$ for
$\Sigma^-\to ne\overline{\nu}$ and $f_1=1.0$, $f_2 = 1.853$, $f_3 = -0.432$, $g_1 = 1.267$, $g_2 = -0.768$, and
$g_3 = 1.765$ in $\Xi^0\to\Sigma^+ e\overline{\nu}$. The anomalous magnetic moments of the baryons contribute to the
RC and we use $\kappa(\Sigma^-) = 0.3764M_N$, $\kappa(\Xi^0) = -0.6661 M_N$, $\kappa(n) =1.9130M_N$, and
$\kappa(\Sigma^+) = 0.8895M_N$, where $M_N$ is the nuclear magneton. These values were extracted from the total
magnetic moments given in Ref.~\cite{part}, using Eq.~(22) of Ref.~\cite{tun91}. The anomalous magnetic moment of the
electron is neglected due to the smallness of its contribution. The values of the masses also come from
Ref.~\cite{part}.

\begingroup
\squeezetable
\begin{table}
\begin{tabular}{lrrrrrrrrrr}
\hline\hline $\sigma$ &
\multicolumn{10}{c}{(a)} \\
\hline
 0.8077 &    0.1218 &    0.1328 &    0.0963 &    0.0421 & $-$0.0159 & $-$0.0681 & $-$0.1058 & $-$0.1212 & $-$0.1063 & $-$0.0506 \\
 0.8056 &    0.1982 &    0.2033 &    0.1443 &    0.0645 & $-$0.0166 & $-$0.0856 & $-$0.1317 & $-$0.1452 & $-$0.1170 & $-$0.0374 \\
 0.8035 &           &    0.2043 &    0.1486 &    0.0698 & $-$0.0107 & $-$0.0785 & $-$0.1223 & $-$0.1322 & $-$0.0991 & $-$0.0141 \\
 0.8014 &           &    0.1974 &    0.1476 &    0.0725 & $-$0.0050 & $-$0.0699 & $-$0.1104 & $-$0.1164 & $-$0.0791 &           \\
 0.7993 &           &    0.1871 &    0.1443 &    0.0738 &    0.0003 & $-$0.0609 & $-$0.0976 & $-$0.0996 & $-$0.0582 &           \\
 0.7972 &           &           &    0.1396 &    0.0743 &    0.0051 & $-$0.0519 & $-$0.0844 & $-$0.0822 & $-$0.0366 &           \\
 0.7951 &           &           &    0.1339 &    0.0741 &    0.0097 & $-$0.0429 & $-$0.0710 & $-$0.0643 & $-$0.0143 &           \\
 0.7930 &           &           &    0.1276 &    0.0734 &    0.0138 & $-$0.0340 & $-$0.0573 & $-$0.0458 &           &           \\
 0.7909 &           &           &           &    0.0721 &    0.0176 & $-$0.0251 & $-$0.0432 & $-$0.0263 &           &           \\
 0.7888 &           &           &           &    0.0700 &    0.0209 & $-$0.0163 & $-$0.0284 & $-$0.0051 &           &           \\
 0.7867 &           &           &           &           &    0.0233 & $-$0.0075 & $-$0.0121 &           &           &           \\ \\
        & \multicolumn{10}{c}{(b)} \\ \hline
 0.8077 &    0.1218 &    0.1328 &    0.0963 &    0.0421 & $-$0.0157 & $-$0.0672 & $-$0.1042 & $-$0.1194 & $-$0.1049 & $-$0.0502 \\
 0.8056 &    0.1982 &    0.2033 &    0.1443 &    0.0645 & $-$0.0163 & $-$0.0847 & $-$0.1302 & $-$0.1436 & $-$0.1159 & $-$0.0371 \\
 0.8035 &           &    0.2043 &    0.1486 &    0.0698 & $-$0.0105 & $-$0.0778 & $-$0.1211 & $-$0.1308 & $-$0.0981 & $-$0.0140 \\
 0.8014 &           &    0.1974 &    0.1476 &    0.0725 & $-$0.0049 & $-$0.0692 & $-$0.1092 & $-$0.1152 & $-$0.0783 &           \\
 0.7993 &           &    0.1871 &    0.1443 &    0.0738 &    0.0004 & $-$0.0603 & $-$0.0966 & $-$0.0986 & $-$0.0576 &           \\
 0.7972 &           &           &    0.1396 &    0.0743 &    0.0052 & $-$0.0514 & $-$0.0836 & $-$0.0814 & $-$0.0363 &           \\
 0.7951 &           &           &    0.1339 &    0.0741 &    0.0097 & $-$0.0425 & $-$0.0703 & $-$0.0637 & $-$0.0142 &           \\
 0.7930 &           &           &    0.1275 &    0.0733 &    0.0138 & $-$0.0337 & $-$0.0568 & $-$0.0453 &           &           \\
 0.7909 &           &           &           &    0.0720 &    0.0176 & $-$0.0249 & $-$0.0428 & $-$0.0260 &           &           \\
 0.7888 &           &           &           &    0.0700 &    0.0209 & $-$0.0162 & $-$0.0282 & $-$0.0050 &           &           \\
 0.7867 &           &           &           &           &    0.0232 & $-$0.0074 & $-$0.0120 &           &           &           \\ \\
 \hline
$\delta$& 0.0500 & 0.1500 & 0.2500 & 0.3500 & 0.4500 & 0.5500 & 0.6500 & 0.7500 & 0.8500 & 0.9500 \\ \\
$\sigma^{max}$& 0.8078 & 0.8078 & 0.8078 & 0.8078 & 0.8078 & 0.8078 & 0.8078 & 0.8078 & 0.8078 & 0.8078 \\
$\sigma^{min}$& 0.8043 & 0.7978 & 0.7925 & 0.7884 & 0.7857 & 0.7847 & 0.7854 & 0.7884 & 0.7939 & 0.8023 \\
\hline\hline
\end{tabular}
\caption{Values of $C_A^\prime$ in $\Sigma^- \to n e {\overline \nu}$ decay by (a) integrating it numerically and (b)
evaluating it analytically. $C_A^\prime$ is given in units of $\textrm{GeV}^2$.}
\end{table}
\endgroup

\begingroup
\squeezetable
\begin{table}
\begin{tabular}{lrrrrrrrrrr}
\hline\hline $\sigma$ &
\multicolumn{10}{c}{(a)} \\
\hline
 0.9091 &    0.0083 &    0.0091 &    0.0071 &    0.0040 &    0.0002 & $-$0.0042 & $-$0.0088 & $-$0.0133 & $-$0.0169 & $-$0.0167 \\
 0.9087 &    0.0305 &    0.0473 &    0.0379 &    0.0184 & $-$0.0065 & $-$0.0339 & $-$0.0608 & $-$0.0826 & $-$0.0901 & $-$0.0481 \\
 0.9083 &           &    0.0610 &    0.0511 &    0.0239 & $-$0.0120 & $-$0.0510 & $-$0.0873 & $-$0.1124 & $-$0.1088 & $-$0.0122 \\
 0.9079 &           &    0.0667 &    0.0593 &    0.0279 & $-$0.0146 & $-$0.0602 & $-$0.1005 & $-$0.1230 & $-$0.1025 &           \\
 0.9075 &           &    0.0686 &    0.0652 &    0.0322 & $-$0.0143 & $-$0.0634 & $-$0.1041 & $-$0.1203 & $-$0.0794 &           \\
 0.9070 &           &    0.0683 &    0.0702 &    0.0371 & $-$0.0112 & $-$0.0615 & $-$0.1001 & $-$0.1073 & $-$0.0439 &           \\
 0.9066 &           &           &    0.0749 &    0.0429 & $-$0.0057 & $-$0.0553 & $-$0.0899 & $-$0.0860 &           &           \\
 0.9062 &           &           &    0.0794 &    0.0496 &    0.0019 & $-$0.0457 & $-$0.0744 & $-$0.0578 &           &           \\
 0.9058 &           &           &           &    0.0568 &    0.0110 & $-$0.0332 & $-$0.0547 & $-$0.0240 &           &           \\
 0.9054 &           &           &           &    0.0639 &    0.0208 & $-$0.0189 & $-$0.0316 &           &           &           \\
 0.9050 &           &           &           &           &    0.0293 & $-$0.0045 & $-$0.0068 &           &           &           \\ \\
        & \multicolumn{10}{c}{(b)} \\ \hline
 0.9091 &    0.0083 &    0.0091 &    0.0071 &    0.0040 &    0.0002 & $-$0.0042 & $-$0.0088 & $-$0.0133 & $-$0.0169 & $-$0.0167 \\
 0.9087 &    0.0305 &    0.0473 &    0.0379 &    0.0184 & $-$0.0065 & $-$0.0339 & $-$0.0608 & $-$0.0826 & $-$0.0901 & $-$0.0481 \\
 0.9083 &           &    0.0610 &    0.0511 &    0.0239 & $-$0.0120 & $-$0.0510 & $-$0.0873 & $-$0.1124 & $-$0.1088 & $-$0.0122 \\
 0.9079 &           &    0.0667 &    0.0593 &    0.0279 & $-$0.0146 & $-$0.0602 & $-$0.1005 & $-$0.1230 & $-$0.1025 &           \\
 0.9075 &           &    0.0686 &    0.0652 &    0.0322 & $-$0.0143 & $-$0.0634 & $-$0.1041 & $-$0.1203 & $-$0.0794 &           \\
 0.9070 &           &    0.0683 &    0.0702 &    0.0371 & $-$0.0112 & $-$0.0615 & $-$0.1001 & $-$0.1073 & $-$0.0439 &           \\
 0.9066 &           &           &    0.0749 &    0.0429 & $-$0.0057 & $-$0.0553 & $-$0.0899 & $-$0.0860 &           &           \\
 0.9062 &           &           &    0.0794 &    0.0496 &    0.0019 & $-$0.0457 & $-$0.0744 & $-$0.0578 &           &           \\
 0.9058 &           &           &           &    0.0568 &    0.0110 & $-$0.0332 & $-$0.0547 & $-$0.0240 &           &           \\
 0.9054 &           &           &           &    0.0639 &    0.0208 & $-$0.0189 & $-$0.0316 &           &           &           \\
 0.9050 &           &           &           &           &    0.0293 & $-$0.0045 & $-$0.0068 &           &           &           \\ \\
\hline
$\delta$ & 0.0500 & 0.1500 & 0.2500 & 0.3500 & 0.4500 & 0.5500 & 0.6500 & 0.7500 & 0.8500 & 0.9500 \\ \\
$\sigma^{max}$ & 0.9091 & 0.9091 & 0.9091 & 0.9091 & 0.9091 & 0.9091 & 0.9091 & 0.9091 & 0.9091 & 0.9091 \\
$\sigma^{min}$ & 0.9083 & 0.9070 & 0.9059 & 0.9051 & 0.9047 & 0.9046 & 0.9049 & 0.9055 & 0.9066 & 0.9082 \\
\hline\hline
\end{tabular}
\caption{Values of $C_{NA}^\prime$ in $\Xi^0 \to \Sigma^+ e {\overline \nu}$ decay by (a) integrating numerically and
(b) evaluating the analytic expression. $C_{NA}^\prime$ is given in units of $\textrm{GeV}^2$ and is multiplied by
100.}
\end{table}
\endgroup

The triple numerical integration results are displayed in entries (a) and the analytical results are displayed in
entries (b) of those two tables. The energies $E$ and $E_2  $ enter through $\delta=E/E_m$ and $\sigma=E_2/M_1$.
$E_m$, $\sigma^{max}$, and $\sigma^{min}$ are determined using the boundaries of the three body region given in
Ref.~\cite{mtz01}. Throughout these tables one can appreciate a satisfactory agreement within two decimal places and
the third place being close within rounding effects.

We have performed another comparison with the numerical results of Ref.~\cite{toth}. Again we shall limit ourselves
to a minimum of examples, since these comparisons were also quite satisfactory. For definiteness, we present the case
of $\Xi^-\to \Lambda e\overline{\nu}$. For this comparison we must use the values of the form factors and of the
energies chosen in this reference. Thus $f_1=1.0$, $g_1=0.249$, $f_2=-0.065$, and $g_2=f_3=g_3=0.0$. The results are
displayed in Table III. The entries correspond to the relative correction defined as
$C_R=100(d\Gamma-d\Gamma_0)/d\Gamma_0$, where $d\Gamma$ is given by Eq.~(\ref{dgi}) with $i=C$ and $d\Gamma_0$ is the
uncorrected decay rate. The upper entries (a) were calculated with our analytical expression, using
$\kappa(\Xi^-)=-0.5940M_N$, and the lower ones (b) come from Ref.~\cite{toth}. The agreement is quite reasonable and
the small differences one observes can be explained by the differences in assumptions. In Ref.~\cite{toth} in the
bremsstrahlung RC the baryons were assumed point-like and higher $(\alpha/\pi)(q/M_1)^n$ contributions $(n\geq 2)$
were included. Instead, we used the theorem of Low \cite{low} and kept only the model independent terms.

Apart from illustration purposes, Tables I-III provide the user of our RC results with numbers to compare with.

\begingroup
\squeezetable
\begin{table}
\begin{tabular}{lrrrrrrrrrr}
\hline\hline $\sigma$ &
\multicolumn{10}{c}{(a)} \\
\hline
0.8558 & 12.4 & 4.6 & 2.5 & 1.1 & 0.0 & $-$1.0 & $-$2.1 & $-$3.4 & $-$5.1 & $-$8.5 \\
0.8546 & 60.4 & 6.2 & 3.3 & 1.7 & 0.4 & $-$0.7 & $-$1.9 & $-$3.3 & $-$5.2 & $-$10.0 \\
0.8534 &  & 8.1 & 3.8 & 1.9 & 0.5 & $-$0.7 & $-$2.0 & $-$3.5 & $-$5.6 & \\
0.8522 &  & 12.5 & 4.4 & 2.1 & 0.6 & $-$0.7 & $-$2.1 & $-$3.7 & $-$6.2 &  \\
0.8510 &  & 42.9 & 5.3 & 2.3 & 0.6 & $-$0.8 & $-$2.3 & $-$4.0 & $-$7.1 &  \\
0.8498 &  &  & 7.2 & 2.6 & 0.6 & $-$0.9 & $-$2.5 & $-$4.4 & $-$10.1 &  \\
0.8485 &  &  & 13.0 & 3.1 & 0.7 & $-$1.0 & $-$2.8 & $-$5.1 &  &  \\
0.8473 &  &  &  & 4.2 & 0.8 & $-$1.2 & $-$3.2 & $-$6.6 &  &  \\
0.8461 &  &  &  & 7.6 & 0.9 & $-$1.5 & $-$4.0 &  &  &  \\
0.8449 &  &  &  &  & 1.3 & $-$2.3 & $-$12.4 &  &  &  \\ \\
       & \multicolumn{10}{c}{(b)} \\ \hline
0.8558 & 12.4 & 4.6 & 2.5 & 1.1 & 0.0 & $-$1.0 & $-$2.1 & $-$3.4 & $-$5.1 & $-$8.5 \\
0.8546 & 60.5 & 6.2 & 3.3 & 1.7 & 0.4 & $-$0.8 & $-$1.9 & $-$3.3 & $-$5.3 & $-$10.0 \\
0.8534 &  & 8.1 & 3.8 & 1.9 & 0.5 & $-$0.7 & $-$2.0 & $-$3.5 & $-$5.6 &  \\
0.8522 &  & 12.6 & 4.4 & 2.1 & 0.6 & $-$0.8 & $-$2.1 & $-$3.7 & $-$6.2 &  \\
0.8510 &  & 42.9 & 5.3 & 2.3 & 0.6 & $-$0.8 & $-$2.3 & $-$4.0 & $-$7.1 &  \\
0.8498 &  &  & 7.2 & 2.6 & 0.6 & $-$0.9 & $-$2.5 & $-$4.4 & $-$10.1 &  \\
0.8485 &  &  & 13.0 & 3.1 & 0.7 & $-$1.0 & $-$2.8 & $-$5.1 &  &  \\
0.8473 &  &  &  & 4.2 & 0.8 & $-$1.2 & $-$3.2 & $-$6.6 &  &  \\
0.8461 &  &  &  & 7.6 & 0.9 & $-$1.5 & $-$4.0 &  &  &  \\
0.8449 &  &  &  &  & 1.3 & $-$2.3 & $-$12.5 &  &  &  \\ \\
\hline
$\delta$ & 0.05 & 0.15 & 0.25 & 0.35 & 0.45 & 0.55 & 0.65 & 0.75 & 0.85 & 0.95 \\ \\
$\sigma^{max}$ & 0.8565 & 0.8565 & 0.8565 & 0.8565 & 0.8565 & 0.8565 & 0.8565 & 0.8565 & 0.8565 & 0.8565 \\
$\sigma^{min}$ & 0.8545 & 0.8510 & 0.8482 & 0.8461 & 0.8448 & 0.8444 & 0.8450 & 0.8466 & 0.8495 & 0.8538 \\
\hline \hline
\end{tabular}
\caption{Values of $C_{R}$ in $\Xi^-\to \Lambda e\overline{\nu} $ decay. (a) corresponds to our analytical expression
and (b) corresponds to Ref.~\cite{toth}.}
\end{table}
\endgroup

\section{Numerical form of the radiative corrections}

As we have seen in the last section, the numerical RC we have discussed are committed to fixing the several form
factors at prescribed values. This is highly undesirable in an experimental analysis. Clearly on the one hand, in the
minimization of such analysis, it is necessary to allow the form factors to be varied freely and, on the other hand,
it is convenient to use the RC in a numerical form. In this section we shall discuss a procedure to obtain numerical
RC that are not committed to fixed values of the form factors and whose use in a simulation amounts to a form
analogous to matrix multiplication.

The RC in Eq.~(\ref{dgi}) are quadratic functions of the form factors, so in general they can be expressed as
\begin{equation}
\Theta_m = \sum_{i\leq j=1}^6 a_{ij}^m f_if_j. \label{tetam}
\end{equation}
In this equation we have momentarily changed our notation and redefined $g_1=f_4$, $g_2=f_5$, and $g_3=f_6$; the
restriction $i\leq j$ reduces the sum to 21 terms. The coefficients $a_{ij}^m$ are functions of $E$ and $E_2$ over
the Dalitz plot. The index $m$ takes the values $m=CI,NI$.

One can calculate the coefficients $a_{ij}^m$ at fixed points $(E,E_2)$ using the analytical results of
Refs.~\cite{tun91} and \cite{tun93} and allow such points to cover a lattice over the Dalitz plot. This lattice
should match at least the bins defined in the experimental arrangement, although for precision RC it should probably
be made larger so as to allow several $(E,E_2)$ points within each bin.

To calculate the coefficients $a_{ij}^m$ it is not necessary to rearrange our final results, either analytical or to
be integrated, so that they take the form (\ref{tetam}). One can calculate them following a systematic procedure. One
chooses fixed $(E,E_2)$ points. Then one fixes $f_1=1$ and $f_i=0$, $i\neq 1$ and obtains $a_{11}^m$; one repeats
this calculation for $f_2=1$ and $f_i=0$, $i\neq 2$ to obtain $a_{22}^m$, and again until $f_6=1$, $f_i=0$,
$i\neq 6$ and $a_{66}^m$ are obtained. Next, one repeats the calculation with $f_1=1$, $f_2=1$, $f_i=0$ $i\neq 1,2$
and from this result one subtracts $a_{11}^m$ and $a_{22}^m$; this way one obtains the coefficient $a_{12}^m$. One
repeats this last step changing $i$ and $j$ until all the interference coefficients $a_{ij}^m$, $i\neq j$, have been
calculated.

To illustrate all this and to further discuss it we have produced arrays presented in two tables, selecting in each
one ten points $(E,E_2)$ over the Dalitz plot. We have chosen two examples, $\Lambda_c^+\to \Lambda e^+\nu$ of a CDB
case which is displayed in Table IV and $\Xi^0 \to \Sigma^+ e\overline{\nu}$ of a NDB case \footnote{The choice of this
decay is of current interest because it is being measured with high statistics by the NA48 Collaboration.}
which is displayed in
Table V. The former also serves as an example of a heavy quark decay. As in the previous section, the more important
purpose is to provide the user with numbers to compare with. The arrays of these two tables were obtained using the
RC in the analytical form. In order to help the interested reader reproduce the entries of Tables IV and V, the
$(\delta,\sigma)$ points were chosen such that $\sigma=M_2/M_1+j \Delta \sigma$, with $j=1,2,\ldots,n$ and
$\Delta \sigma=(E_2^T-M_2)/(nM_1)$, with $E_2^T=(1/2)\{(M_1-m)^2 + M_2^2\}/(M_1-m)$. Besides, in the $\Lambda_c^+$ case
we used the formulas for the charged assignments $A^-$, $B^0$, $l^-$ of the CDB case of the previous sections and then
applied the rules of Ref.~\cite{rfm02} to obtain the results for the charge assignments $A^+$, $B^0$, $l^+$ of this
particular case. In these tables we have restored our standard notation for the axial-vector form factors $g_1$, $g_2$,
and $g_3$. The masses used are those of Sec.~III, $M_{\Lambda_c^+}$ comes from Ref.~\cite{part}, and we assume an
estimate for $\kappa(\Lambda_c^+)=0.1106M_N$.

The first fact that appears in these Tables IV and V is that the RC do not depend on the form factor products
$f_1g_3$, $f_2g_3$, $f_3g_1$, $f_3g_2$, and $f_3g_3$. The nonappearance of these products cannot be seen
easily in our final results of Sec.~II. The other fact is that the nonzero RC to each form factor product vary
appreciably from one $(E,E_2)$ point to another. This means that replacing the precision results of Sec.~II with an
array of only a few columns over the Dalitz plot is far from satisfactory. Therefore the lattice of $(E,E_2)$ points
must be much finer than only a few points.

The use of this third presentation of RC is very practical in the sense that such RC can be calculated separately and
only the arrays should be feed into the Monte Carlo simulation.

\begin{turnpage}

\begingroup
\squeezetable
\begin{table}
\begin{center}
\begin{tabular}{
l r@{.}l@{}r r@{.}l@{}r r@{.}l@{}r r@{.}l@{}r r@{.}l@{}r
r@{.}l@{}r r@{.}l@{}r r@{.}l@{}r r@{.}l@{}r r@{.}l@{}r}
\hline\hline &
\multicolumn{3}{c}{(0.15,0.5995)} &
\multicolumn{3}{c}{(0.45,0.5995)} &
\multicolumn{3}{c}{(0.75,0.5995)} &
\multicolumn{3}{c}{(0.95,0.5995)} &
\multicolumn{3}{c}{(0.25,0.5602)} &
\multicolumn{3}{c}{(0.55,0.5602)} &
\multicolumn{3}{c}{(0.85,0.5602)} &
\multicolumn{3}{c}{(0.45,0.5210)} &
\multicolumn{3}{c}{(0.75,0.5210)} &
\multicolumn{3}{c}{(0.65,0.4948)}  \\ \hline
$f_1^2 $ & $ 4$ & 639 & \multicolumn{1}{c}{} & $ 5$ & 764 & $\times 10^{-1}$ & $-4$ & 632 & \multicolumn{1}{c}{} & $-2$ & 647 & \multicolumn{1}{c}{} & $ 3$ & 369 & \multicolumn{1}{c}{} & $-7$ & 737 & $\times 10^{-1}$ & $-3$ & 403 & \multicolumn{1}{c}{} & $ 1$ & 176 & \multicolumn{1}{c}{} & $-2$ & 016 & \multicolumn{1}{c}{} & $-3$ & 654 & $\times 10^{-1}$ \\
$f_2^2 $ & $ 6$ & 914 & $\times 10^{-1}$ & $ 1$ & 007 & $\times 10^{-1}$ & $-6$ & 431 & $\times 10^{-1}$ & $-7$ & 604 & $\times 10^{-1}$ & $ 5$ & 006 & $\times 10^{-1}$ & $-6$ & 431 & $\times 10^{-2}$ & $-7$ & 948 & $\times 10^{-1}$ & $ 8$ & 519 & $\times 10^{-2}$ & $-3$ & 917 & $\times 10^{-1}$ & $-7$ & 668 & $\times 10^{-2}$ \\
$f_3^2 $ & $-2$ & 944 & $\times 10^{-1}$ & $-2$ & 499 & $\times 10^{-2}$ & $ 3$ & 269 & $\times 10^{-1}$ & $-6$ & 177 & $\times 10^{-2}$ & $-1$ & 813 & $\times 10^{-1}$ & $ 7$ & 650 & $\times 10^{-2}$ & $ 2$ & 357 & $\times 10^{-2}$ & $-1$ & 037 & $\times 10^{-1}$ & $ 6$ & 046 & $\times 10^{-2}$ & $ 1$ & 093 & $\times 10^{-2}$ \\
$g_1^2 $ & $ 5$ & 963 & \multicolumn{1}{c}{} & $ 8$ & 010 & $\times 10^{-1}$ & $-5$ & 674 & \multicolumn{1}{c}{} & $-6$ & 435 & \multicolumn{1}{c}{} & $ 6$ & 416 & \multicolumn{1}{c}{} & $-1$ & 008 & \multicolumn{1}{c}{} & $-1$ & 013 & $\times 10^{+1}$ & $ 2$ & 271 & \multicolumn{1}{c}{} & $-1$ & 038 & $\times 10^{+1}$ & $-9$ & 480 & \multicolumn{1}{c}{} \\
$g_2^2 $ & $ 4$ & 804 & $\times 10^{-1}$ & $ 8$ & 761 & $\times 10^{-2}$ & $-3$ & 905 & $\times 10^{-1}$ & $-9$ & 652 & $\times 10^{-1}$ & $ 7$ & 038 & $\times 10^{-1}$ & $-1$ & 568 & $\times 10^{-2}$ & $-1$ & 565 & \multicolumn{1}{c}{} & $ 2$ & 244 & $\times 10^{-1}$ & $-1$ & 971 & \multicolumn{1}{c}{} & $-2$ & 324 & \multicolumn{1}{c}{} \\
$g_3^2 $ & $-3$ & 010 & $\times 10^{-2}$ & $-2$ & 555 & $\times 10^{-3}$ & $ 3$ & 343 & $\times 10^{-2}$ & $-6$ & 316 & $\times 10^{-3}$ & $-1$ & 244 & $\times 10^{-2}$ & $ 5$ & 250 & $\times 10^{-3}$ & $ 1$ & 618 & $\times 10^{-3}$ & $-3$ & 359 & $\times 10^{-3}$ & $ 1$ & 959 & $\times 10^{-3}$ & $ 7$ & 272 & $\times 10^{-5}$ \\
$f_1f_2$ & $ 1$ & 011 & \multicolumn{1}{c}{} & $ 1$ & 398 & $\times 10^{-1}$ & $-9$ & 471 & $\times 10^{-1}$ & $-1$ & 170 & \multicolumn{1}{c}{} & $ 1$ & 055 & \multicolumn{1}{c}{} & $-1$ & 736 & $\times 10^{-1}$ & $-1$ & 517 & \multicolumn{1}{c}{} & $ 3$ & 411 & $\times 10^{-1}$ & $-9$ & 479 & $\times 10^{-1}$ & $-1$ & 988 & $\times 10^{-1}$ \\
$f_1f_3$ & $ 5$ & 303 & $\times 10^{-1}$ & $ 3$ & 216 & $\times 10^{-2}$ & $-6$ & 074 & $\times 10^{-1}$ & $ 1$ & 133 & $\times 10^{-1}$ & $ 3$ & 281 & $\times 10^{-1}$ & $-1$ & 603 & $\times 10^{-1}$ & $-4$ & 672 & $\times 10^{-2}$ & $ 1$ & 912 & $\times 10^{-1}$ & $-1$ & 231 & $\times 10^{-1}$ & $-2$ & 662 & $\times 10^{-2}$ \\
$f_2f_3$ & $ 2$ & 176 & $\times 10^{-1}$ & $ 1$ & 947 & $\times 10^{-2}$ & $-2$ & 403 & $\times 10^{-1}$ & $ 4$ & 551 & $\times 10^{-2}$ & $ 1$ & 561 & $\times 10^{-1}$ & $-6$ & 085 & $\times 10^{-2}$ & $-1$ & 937 & $\times 10^{-2}$ & $ 1$ & 038 & $\times 10^{-1}$ & $-5$ & 612 & $\times 10^{-2}$ & $-9$ & 102 & $\times 10^{-3}$ \\
$g_1g_2$ & $-2$ & 084 & \multicolumn{1}{c}{} & $-3$ & 016 & $\times 10^{-1}$ & $ 1$ & 792 & \multicolumn{1}{c}{} & $ 4$ & 204 & \multicolumn{1}{c}{} & $-3$ & 801 & \multicolumn{1}{c}{} & $ 3$ & 736 & $\times 10^{-1}$ & $ 7$ & 428 & \multicolumn{1}{c}{} & $-1$ & 481 & \multicolumn{1}{c}{} & $ 8$ & 939 & \multicolumn{1}{c}{} & $ 9$ & 367 & \multicolumn{1}{c}{} \\
$g_1g_3$ & $-5$ & 523 & $\times 10^{-1}$ & $-5$ & 975 & $\times 10^{-2}$ & $ 5$ & 949 & $\times 10^{-1}$ & $-1$ & 138 & $\times 10^{-1}$ & $-3$ & 635 & $\times 10^{-1}$ & $ 1$ & 315 & $\times 10^{-1}$ & $ 4$ & 321 & $\times 10^{-2}$ & $-2$ & 196 & $\times 10^{-1}$ & $ 1$ & 165 & $\times 10^{-1}$ & $ 1$ & 785 & $\times 10^{-2}$ \\
$g_2g_3$ & $ 2$ & 176 & $\times 10^{-1}$ & $ 1$ & 947 & $\times 10^{-2}$ & $-2$ & 403 & $\times 10^{-1}$ & $ 4$ & 551 & $\times 10^{-2}$ & $ 1$ & 561 & $\times 10^{-1}$ & $-6$ & 085 & $\times 10^{-2}$ & $-1$ & 937 & $\times 10^{-2}$ & $ 1$ & 038 & $\times 10^{-1}$ & $-5$ & 612 & $\times 10^{-2}$ & $-9$ & 102 & $\times 10^{-3}$ \\
$f_1g_1$ & $-6$ & 338 & $\times 10^{-3}$ & $-7$ & 034 & $\times 10^{-1}$ & $-6$ & 223 & $\times 10^{-2}$ & $ 2$ & 432 & \multicolumn{1}{c}{} & $-8$ & 672 & $\times 10^{-1}$ & $-1$ & 921 & \multicolumn{1}{c}{} & $ 2$ & 286 & \multicolumn{1}{c}{} & $-2$ & 952 & \multicolumn{1}{c}{} & $ 3$ & 431 & $\times 10^{-1}$ & $-1$ & 730 & \multicolumn{1}{c}{} \\
$f_1g_2$ & $-6$ & 962 & $\times 10^{-3}$ & $ 3$ & 520 & $\times 10^{-1}$ & $ 2$ & 672 & $\times 10^{-2}$ & $-1$ & 246 & \multicolumn{1}{c}{} & $ 4$ & 037 & $\times 10^{-1}$ & $ 9$ & 504 & $\times 10^{-1}$ & $-1$ & 183 & \multicolumn{1}{c}{} & $ 1$ & 440 & \multicolumn{1}{c}{} & $-2$ & 088 & $\times 10^{-1}$ & $ 8$ & 390 & $\times 10^{-1}$ \\
$f_1g_3$ & $ 0$ & 000 & \multicolumn{1}{c}{} & $ 0$ & 000 & \multicolumn{1}{c}{} & $ 0$ & 000 & \multicolumn{1}{c}{} & $ 0$ & 000 & \multicolumn{1}{c}{} & $ 0$ & 000 & \multicolumn{1}{c}{} & $ 0$ & 000 & \multicolumn{1}{c}{} & $ 0$ & 000 & \multicolumn{1}{c}{} & $ 0$ & 000 & \multicolumn{1}{c}{} & $ 0$ & 000 & \multicolumn{1}{c}{} & $ 0$ & 000 & \multicolumn{1}{c}{} \\
$f_2g_1$ & $-2$ & 744 & $\times 10^{-2}$ & $-1$ & 059 & \multicolumn{1}{c}{} & $-9$ & 851 & $\times 10^{-2}$ & $ 3$ & 618 & \multicolumn{1}{c}{} & $-1$ & 348 & \multicolumn{1}{c}{} & $-2$ & 898 & \multicolumn{1}{c}{} & $ 3$ & 387 & \multicolumn{1}{c}{} & $-4$ & 477 & \multicolumn{1}{c}{} & $ 4$ & 752 & $\times 10^{-1}$ & $-2$ & 625 & \multicolumn{1}{c}{} \\
$f_2g_2$ & $-1$ & 036 & $\times 10^{-2}$ & $ 5$ & 239 & $\times 10^{-1}$ & $ 3$ & 977 & $\times 10^{-2}$ & $-1$ & 855 & \multicolumn{1}{c}{} & $ 6$ & 009 & $\times 10^{-1}$ & $ 1$ & 415 & \multicolumn{1}{c}{} & $-1$ & 761 & \multicolumn{1}{c}{} & $ 2$ & 143 & \multicolumn{1}{c}{} & $-3$ & 107 & $\times 10^{-1}$ & $ 1$ & 249 & \multicolumn{1}{c}{} \\
$f_2g_3$ & $ 0$ & 000 & \multicolumn{1}{c}{} & $ 0$ & 000 & \multicolumn{1}{c}{} & $ 0$ & 000 & \multicolumn{1}{c}{} & $ 0$ & 000 & \multicolumn{1}{c}{} & $ 0$ & 000 & \multicolumn{1}{c}{} & $ 0$ & 000 & \multicolumn{1}{c}{} & $ 0$ & 000 & \multicolumn{1}{c}{} & $ 0$ & 000 & \multicolumn{1}{c}{} & $ 0$ & 000 & \multicolumn{1}{c}{} & $ 0$ & 000 & \multicolumn{1}{c}{} \\
$f_3g_1$ & $ 0$ & 000 & \multicolumn{1}{c}{} & $ 0$ & 000 & \multicolumn{1}{c}{} & $ 0$ & 000 & \multicolumn{1}{c}{} & $ 0$ & 000 & \multicolumn{1}{c}{} & $ 0$ & 000 & \multicolumn{1}{c}{} & $ 0$ & 000 & \multicolumn{1}{c}{} & $ 0$ & 000 & \multicolumn{1}{c}{} & $ 0$ & 000 & \multicolumn{1}{c}{} & $ 0$ & 000 & \multicolumn{1}{c}{} & $ 0$ & 000 & \multicolumn{1}{c}{} \\
$f_3g_2$ & $ 0$ & 000 & \multicolumn{1}{c}{} & $ 0$ & 000 & \multicolumn{1}{c}{} & $ 0$ & 000 & \multicolumn{1}{c}{} & $ 0$ & 000 & \multicolumn{1}{c}{} & $ 0$ & 000 & \multicolumn{1}{c}{} & $ 0$ & 000 & \multicolumn{1}{c}{} & $ 0$ & 000 & \multicolumn{1}{c}{} & $ 0$ & 000 & \multicolumn{1}{c}{} & $ 0$ & 000 & \multicolumn{1}{c}{} & $ 0$ & 000 & \multicolumn{1}{c}{} \\
$f_3g_3$ & $ 0$ & 000 & \multicolumn{1}{c}{} & $ 0$ & 000 & \multicolumn{1}{c}{} & $ 0$ & 000 & \multicolumn{1}{c}{} & $ 0$ & 000 & \multicolumn{1}{c}{} & $ 0$ & 000 & \multicolumn{1}{c}{} & $ 0$ & 000 & \multicolumn{1}{c}{} & $ 0$ & 000 & \multicolumn{1}{c}{} & $ 0$ & 000 & \multicolumn{1}{c}{} & $ 0$ & 000 & \multicolumn{1}{c}{} & $ 0$ & 000 & \multicolumn{1}{c}{} \\
\hline
\end{tabular}
\caption{Numerical arrays of the coefficients $a_{ij}^{CI}$ in $\textrm{GeV}^2$ of Eq.~(8) evaluated at ten points
$(\delta,\sigma)$ (headings of columns) over the unpolarized Dalitz plot of
$\Lambda_c^+ \to \Lambda e^+ \nu$ decay.}
\end{center}
\end{table}
\endgroup

\end{turnpage}

\begin{turnpage}

\begingroup
\squeezetable
\begin{table}
\begin{center}
\begin{tabular}{
l r@{.}l@{}r r@{.}l@{}r r@{.}l@{}r r@{.}l@{}r r@{.}l@{}r
r@{.}l@{}r r@{.}l@{}r r@{.}l@{}r r@{.}l@{}r r@{.}l@{}r }
\hline\hline &
\multicolumn{3}{c}{(0.05,0.9086)} &
\multicolumn{3}{c}{(0.35,0.9086)} &
\multicolumn{3}{c}{(0.65,0.9086)} &
\multicolumn{3}{c}{(0.95,0.9086)} &
\multicolumn{3}{c}{(0.25,0.9073)} &
\multicolumn{3}{c}{(0.55,0.9073)} &
\multicolumn{3}{c}{(0.75,0.9073)} &
\multicolumn{3}{c}{(0.45,0.9059)} &
\multicolumn{3}{c}{(0.65,0.9059)} &
\multicolumn{3}{c}{(0.55,0.9050)}
\\ \hline
$f_1^2 $ & $ 4$ & 874 & $\times 10^{-2}$ & $ 9$ & 845 & $\times 10^{-2}$ & $ 1$ & 783 & $\times 10^{-2}$ & $-1$ & 968 & $\times 10^{-2}$ & $ 6$ & 750 & $\times 10^{-2}$ & $ 2$ & 728 & $\times 10^{-2}$ & $-1$ & 885 & $\times 10^{-2}$ & $ 2$ & 339 & $\times 10^{-2}$ & $-8$ & 097 & $\times 10^{-3}$ & $-1$ & 071 & $\times 10^{-3}$\\
$f_2^2 $ & $ 5$ & 047 & $\times 10^{-4}$ & $ 8$ & 128 & $\times 10^{-4}$ & $ 1$ & 708 & $\times 10^{-4}$ & $-2$ & 698 & $\times 10^{-4}$ & $ 6$ & 142 & $\times 10^{-4}$ & $ 2$ & 818 & $\times 10^{-4}$ & $-9$ & 249 & $\times 10^{-5}$ & $ 2$ & 133 & $\times 10^{-4}$ & $ 2$ & 666 & $\times 10^{-6}$ & $ 1$ & 750 & $\times 10^{-5}$\\
$f_3^2 $ & $-2$ & 270 & $\times 10^{-4}$ & $-6$ & 500 & $\times 10^{-4}$ & $-9$ & 241 & $\times 10^{-5}$ & $ 3$ & 358 & $\times 10^{-5}$ & $-2$ & 443 & $\times 10^{-4}$ & $-6$ & 334 & $\times 10^{-5}$ & $ 1$ & 487 & $\times 10^{-4}$ & $-3$ & 114 & $\times 10^{-5}$ & $ 8$ & 899 & $\times 10^{-5}$ & $ 2$ & 953 & $\times 10^{-5}$\\
$g_1^2 $ & $ 8$ & 473 & $\times 10^{-2}$ & $ 1$ & 201 & $\times 10^{-1}$ & $ 2$ & 695 & $\times 10^{-2}$ & $-5$ & 073 & $\times 10^{-2}$ & $ 1$ & 718 & $\times 10^{-1}$ & $ 8$ & 107 & $\times 10^{-2}$ & $-1$ & 978 & $\times 10^{-2}$ & $ 1$ & 378 & $\times 10^{-1}$ & $ 7$ & 920 & $\times 10^{-3}$ & $ 3$ & 924 & $\times 10^{-2}$\\
$g_2^2 $ & $ 3$ & 134 & $\times 10^{-4}$ & $ 1$ & 856 & $\times 10^{-4}$ & $ 8$ & 773 & $\times 10^{-5}$ & $-2$ & 664 & $\times 10^{-4}$ & $ 7$ & 566 & $\times 10^{-4}$ & $ 4$ & 189 & $\times 10^{-4}$ & $ 5$ & 428 & $\times 10^{-5}$ & $ 9$ & 168 & $\times 10^{-4}$ & $ 1$ & 964 & $\times 10^{-4}$ & $ 3$ & 790 & $\times 10^{-4}$\\
$g_3^2 $ & $-5$ & 090 & $\times 10^{-7}$ & $-1$ & 457 & $\times 10^{-6}$ & $-2$ & 072 & $\times 10^{-7}$ & $ 7$ & 528 & $\times 10^{-8}$ & $-3$ & 654 & $\times 10^{-7}$ & $-9$ & 475 & $\times 10^{-8}$ & $ 2$ & 225 & $\times 10^{-7}$ & $-2$ & 331 & $\times 10^{-8}$ & $ 6$ & 661 & $\times 10^{-8}$ & $ 7$ & 372 & $\times 10^{-9}$\\
$f_1f_2$ & $ 5$ & 638 & $\times 10^{-4}$ & $ 9$ & 122 & $\times 10^{-4}$ & $ 1$ & 875 & $\times 10^{-4}$ & $-3$ & 016 & $\times 10^{-4}$ & $ 8$ & 794 & $\times 10^{-4}$ & $ 3$ & 890 & $\times 10^{-4}$ & $-1$ & 641 & $\times 10^{-4}$ & $ 3$ & 700 & $\times 10^{-4}$ & $-4$ & 900 & $\times 10^{-5}$ & $ 8$ & 182 & $\times 10^{-6}$\\
$f_1f_3$ & $ 2$ & 474 & $\times 10^{-4}$ & $ 7$ & 098 & $\times 10^{-4}$ & $ 9$ & 664 & $\times 10^{-5}$ & $-3$ & 723 & $\times 10^{-5}$ & $ 2$ & 585 & $\times 10^{-4}$ & $ 6$ & 115 & $\times 10^{-5}$ & $-1$ & 671 & $\times 10^{-4}$ & $ 2$ & 655 & $\times 10^{-5}$ & $-1$ & 010 & $\times 10^{-4}$ & $-3$ & 519 & $\times 10^{-5}$\\
$f_2f_3$ & $ 2$ & 291 & $\times 10^{-5}$ & $ 6$ & 557 & $\times 10^{-5}$ & $ 9$ & 365 & $\times 10^{-6}$ & $-3$ & 382 & $\times 10^{-6}$ & $ 2$ & 544 & $\times 10^{-5}$ & $ 6$ & 820 & $\times 10^{-6}$ & $-1$ & 511 & $\times 10^{-5}$ & $ 3$ & 760 & $\times 10^{-6}$ & $-9$ & 076 & $\times 10^{-6}$ & $-2$ & 900 & $\times 10^{-6}$\\
$g_1g_2$ & $-7$ & 212 & $\times 10^{-3}$ & $-4$ & 911 & $\times 10^{-3}$ & $-1$ & 891 & $\times 10^{-3}$ & $ 5$ & 952 & $\times 10^{-3}$ & $-2$ & 032 & $\times 10^{-2}$ & $-1$ & 045 & $\times 10^{-2}$ & $ 2$ & 576 & $\times 10^{-4}$ & $-2$ & 196 & $\times 10^{-2}$ & $-3$ & 032 & $\times 10^{-3}$ & $-7$ & 708 & $\times 10^{-3}$\\
$g_1g_3$ & $-2$ & 541 & $\times 10^{-4}$ & $-7$ & 244 & $\times 10^{-4}$ & $-1$ & 072 & $\times 10^{-4}$ & $ 3$ & 692 & $\times 10^{-5}$ & $-2$ & 811 & $\times 10^{-4}$ & $-7$ & 867 & $\times 10^{-5}$ & $ 1$ & 614 & $\times 10^{-4}$ & $-4$ & 222 & $\times 10^{-5}$ & $ 9$ & 575 & $\times 10^{-5}$ & $ 3$ & 032 & $\times 10^{-5}$\\
$g_2g_3$ & $ 2$ & 291 & $\times 10^{-5}$ & $ 6$ & 557 & $\times 10^{-5}$ & $ 9$ & 365 & $\times 10^{-6}$ & $-3$ & 382 & $\times 10^{-6}$ & $ 2$ & 544 & $\times 10^{-5}$ & $ 6$ & 820 & $\times 10^{-6}$ & $-1$ & 511 & $\times 10^{-5}$ & $ 3$ & 760 & $\times 10^{-6}$ & $-9$ & 076 & $\times 10^{-6}$ & $-2$ & 900 & $\times 10^{-6}$\\
$f_1g_1$ & $-2$ & 678 & $\times 10^{-3}$ & $ 1$ & 609 & $\times 10^{-4}$ & $ 9$ & 352 & $\times 10^{-4}$ & $-2$ & 700 & $\times 10^{-3}$ & $-2$ & 051 & $\times 10^{-3}$ & $ 3$ & 242 & $\times 10^{-3}$ & $ 1$ & 489 & $\times 10^{-3}$ & $ 3$ & 140 & $\times 10^{-3}$ & $ 2$ & 945 & $\times 10^{-3}$ & $ 3$ & 124 & $\times 10^{-3}$\\
$f_1g_2$ & $ 2$ & 484 & $\times 10^{-4}$ & $-1$ & 926 & $\times 10^{-5}$ & $-8$ & 964 & $\times 10^{-5}$ & $ 2$ & 580 & $\times 10^{-4}$ & $ 1$ & 858 & $\times 10^{-4}$ & $-3$ & 063 & $\times 10^{-4}$ & $-1$ & 364 & $\times 10^{-4}$ & $-2$ & 898 & $\times 10^{-4}$ & $-2$ & 690 & $\times 10^{-4}$ & $-2$ & 829 & $\times 10^{-4}$\\
$f_1g_3$ & $ 0$ & 000 & \multicolumn{1}{c}{} & $ 0$ & 000 & \multicolumn{1}{c}{} & $ 0$ & 000 & \multicolumn{1}{c}{} & $ 0$ & 000 & \multicolumn{1}{c}{} & $ 0$ & 000 & \multicolumn{1}{c}{} & $ 0$ & 000 & \multicolumn{1}{c}{} & $ 0$ & 000 & \multicolumn{1}{c}{} & $ 0$ & 000 & \multicolumn{1}{c}{} & $ 0$ & 000 & \multicolumn{1}{c}{} & $ 0$ & 000 & \multicolumn{1}{c}{}\\
$f_2g_1$ & $-4$ & 874 & $\times 10^{-3}$ & $ 4$ & 487 & $\times 10^{-4}$ & $ 1$ & 824 & $\times 10^{-3}$ & $-5$ & 142 & $\times 10^{-3}$ & $-3$ & 363 & $\times 10^{-3}$ & $ 6$ & 360 & $\times 10^{-3}$ & $ 2$ & 878 & $\times 10^{-3}$ & $ 6$ & 339 & $\times 10^{-3}$ & $ 5$ & 704 & $\times 10^{-3}$ & $ 6$ & 106 & $\times 10^{-3}$\\
$f_2g_2$ & $ 4$ & 732 & $\times 10^{-4}$ & $-3$ & 667 & $\times 10^{-5}$ & $-1$ & 707 & $\times 10^{-4}$ & $ 4$ & 914 & $\times 10^{-4}$ & $ 3$ & 539 & $\times 10^{-4}$ & $-5$ & 834 & $\times 10^{-4}$ & $-2$ & 598 & $\times 10^{-4}$ & $-5$ & 520 & $\times 10^{-4}$ & $-5$ & 124 & $\times 10^{-4}$ & $-5$ & 388 & $\times 10^{-4}$\\
$f_2g_3$ & $ 0$ & 000 & \multicolumn{1}{c}{} & $ 0$ & 000 & \multicolumn{1}{c}{} & $ 0$ & 000 & \multicolumn{1}{c}{} & $ 0$ & 000 & \multicolumn{1}{c}{} & $ 0$ & 000 & \multicolumn{1}{c}{} & $ 0$ & 000 & \multicolumn{1}{c}{} & $ 0$ & 000 & \multicolumn{1}{c}{} & $ 0$ & 000 & \multicolumn{1}{c}{} & $ 0$ & 000 & \multicolumn{1}{c}{} & $ 0$ & 000 & \multicolumn{1}{c}{}\\
$f_3g_1$ & $ 0$ & 000 & \multicolumn{1}{c}{} & $ 0$ & 000 & \multicolumn{1}{c}{} & $ 0$ & 000 & \multicolumn{1}{c}{} & $ 0$ & 000 & \multicolumn{1}{c}{} & $ 0$ & 000 & \multicolumn{1}{c}{} & $ 0$ & 000 & \multicolumn{1}{c}{} & $ 0$ & 000 & \multicolumn{1}{c}{} & $ 0$ & 000 & \multicolumn{1}{c}{} & $ 0$ & 000 & \multicolumn{1}{c}{} & $ 0$ & 000 & \multicolumn{1}{c}{}\\
$f_3g_2$ & $ 0$ & 000 & \multicolumn{1}{c}{} & $ 0$ & 000 & \multicolumn{1}{c}{} & $ 0$ & 000 & \multicolumn{1}{c}{} & $ 0$ & 000 & \multicolumn{1}{c}{} & $ 0$ & 000 & \multicolumn{1}{c}{} & $ 0$ & 000 & \multicolumn{1}{c}{} & $ 0$ & 000 & \multicolumn{1}{c}{} & $ 0$ & 000 & \multicolumn{1}{c}{} & $ 0$ & 000 & \multicolumn{1}{c}{} & $ 0$ & 000 & \multicolumn{1}{c}{}\\
$f_3g_3$ & $ 0$ & 000 & \multicolumn{1}{c}{} & $ 0$ & 000 & \multicolumn{1}{c}{} & $ 0$ & 000 & \multicolumn{1}{c}{} & $ 0$ & 000 & \multicolumn{1}{c}{} & $ 0$ & 000 & \multicolumn{1}{c}{} & $ 0$ & 000 & \multicolumn{1}{c}{} & $ 0$ & 000 & \multicolumn{1}{c}{} & $ 0$ & 000 & \multicolumn{1}{c}{} & $ 0$ & 000 & \multicolumn{1}{c}{} & $ 0$ & 000 & \multicolumn{1}{c}{}\\
\hline
\end{tabular}
\caption{Numerical arrays of the coefficients $a_{ij}^{NI}$ in $\textrm{GeV}^2$ of Eq.~(8) evaluated at ten points
$(\delta,\sigma)$ (headings of columns) over the unpolarized Dalitz plot of $\Xi^0 \to \Sigma^+ e \nu$ decay.}
\end{center}
\end{table}
\endgroup

\end{turnpage}

\section{Summary and conclusions}

In this paper we have proposed and discussed a numerical form of the model independent RC to baryon semileptonic
decays which is not compromised to fixing the values of the form factors at prescribed values. This last would not
be acceptable in decays involving heavy quarks. The numerical RC are organized in arrays. Each column of the array is
calculated at fixed $(E,E_2)$ and applied as a matrix multiplication to the form factor products. The $(E,E_2)$ points
cover a lattice over the Dalitz plot.

The inclusion of precision RC in an experimental analysis is reduced to simple multiplication, avoiding the calculation of the
long analytical expressions as $E$ and $E_2$ are varied. However, in a precision experiment possibly involving 150,
200, and even 300 bins over the Dalitz plot the number of columns in the RC arrays should be at least just as many. It
may be required that several columns be reproduced in finer subdivisions within each bin, possibly four, eight, or
even more. For example, one may require that the numerical changes of the $a_{ij}^m$ coefficients between neighboring
$(E,E_2)$ points do not exceed two decimal places within rounding of the third decimal place.

To close, let us recall that our results are general within our approximation. They can be applied in the other four
charge assignments of baryons involving heavy quarks and whether the charged lepton is $e^\pm$, $\mu^\pm$, or
$\tau^\pm$. They are model independent and not compromised to fixing the form factor at determined values. Let us
finally emphasize that Tables I-V should serve for comparison purposes and help the user to check his/her use of our
analytical results.

\acknowledgments
The authors are grateful to Consejo Nacional de Ciencia y Tecnolog{\'\i}a (Mexico) for partial support. J.J.~Torres,
A.~Mart{\'\i}nez, and M.~Neri were partially supported by Comisi\'on de Operaci\'on y Fomento de Actividades
Acad\'emicas (Instituto Polit\'ecnico Nacional). They also wish to thank the warm hospitality extended to them at
IF-UASLP, where part of this paper was performed. R.~F.-M.\ was also partially supported by Fondo de Apoyo a la
Investigaci\'on (Universidad Aut\'onoma de San Luis Potos{\'\i}).

\appendix

\section{Collection of the $H_i^\prime$ and $\rho_i$ functions}

In this section we present the $H_i^\prime$ and the $\rho _i$ functions of $C_A^\prime$ in Eq.~(\ref{CAp}) and
$C_{NA}^\prime$ in Eq.~(\ref{CNAp}), respectively. They read
\begin{equation*}
H_0^\prime = p_2l\frac{Q_3+Q_4E_\nu^0}{2}-\frac{p_2lE}{M_1} \left[ \frac{B^+}{2}+A^+\right],
\end{equation*}
\begin{equation*}
H_2 ^\prime = \frac{p_2l(1-\beta^2)}{2}\left[-E_\nu^0(Q_1+Q_3)+Q_4(E+E_\nu^0)^2+(Q_2+Q_4)p_2ly_0+Q_2 p_2^2\right],
\end{equation*}
\begin{eqnarray*}
H_3^\prime & = & \frac{p_2l}{2} \left\{(Q_1+Q_3)\left[E_\nu^0-E\frac{1+\beta^2}{2}\right] - (E+E_\nu^0+p_2\beta y_0)
Q_3 \right. \\
&  & \mbox{} -Q_2 \left. \left[p_2ly_0+p_2^2-E(E+E_\nu^0)\frac{1+\beta^2}{2}\right] + Q_4\left[E_\nu^0\beta p_2y_0
- E(E+E_\nu^0) \frac{1-3\beta^2}{2}\right] \right\} \\
&  & \mbox{} +\left. \frac{p_2lE(1-\beta^2)}{2M_1}\right\{ \left[ (E+E_\nu^0)B^--EB^+\right] + 2E_\nu^0[A^-+g_1
(f_2-g_2  )] \\
&  & \mbox{} - \left. 4Eg_1(f_2+g_2)+2M_1\frac{h^+}{e}(2E+E_\nu^0)\right\},
\end{eqnarray*}
\begin{eqnarray*}
H_4^\prime & = & \frac{p_2l}{4}\left\{Q_1E+Q_3(2E_\nu^0+3E)-(Q_2+Q_4)E[E(1+\beta^2)+E_\nu^0]\right. \\
&  & \mbox{} + \left. 2Q_4[E_\nu^0(E_\nu^0-E)-p_2ly_0] \right\} + \frac{p_2lE}{2M_1}\left[ -(E+E_\nu^0)B^-+EB^++p_2
\beta y_0C^-\right] \\
&  & \mbox{} +\ frac{p_2lE}{M_1}\left[ -M_1\frac{h^+}{e}[E_\nu^0+2E(1-\beta^2)]+M_1\frac{h^-}{e}\beta^2E-E_\nu^0
(f_1f_2+g_1g_2)(1+\beta x_0) \right. \\
&  & \mbox{} + \left. E_\nu^0g_1(f_2+3g_2)+E\left[ (1-\beta^2)(f_1f_2+4g_1g_2-g_1f_2)-f_1f_2+g_1(3f_2-2g_2)\right]
\right],
\end{eqnarray*}
\begin{eqnarray*}
H_5^\prime & = & \frac{p_2l^2}{4} \left\{Q_1+3Q_3-Q_2(E+E_\nu^0)-Q_4(3E+7E_\nu^0)+\frac{2}{M_1}\left[ (E_\nu^0-E)
B^-+EB^++E_\nu^0C^-\right] \right\}  \\
&  & \mbox{} + p_2l^2\left[ -\frac{h^+}{e}(2E+E_\nu^0)-\frac{h^-}{e} 2E_\nu^0+\frac{E_\nu^0}{M_1}(2f_1f_2+3g_1g_2
-g_1f_2  ) + \frac{2E}{M_1}g_1(f_2+g_2) \right],
\end{eqnarray*}
\begin{equation*}
H_6^\prime = \frac{p_2l(1-\beta^2)}{4}\left[Q_1+Q_3-(Q_2  +Q_4)(E+E_\nu^0)\right],
\end{equation*}
\begin{eqnarray*}
H_7^\prime & = &  -\frac{p_2l}{4}\left[ (Q_1+Q_3)\frac{2E-E_\nu^0}{E}+(Q_2+Q_4)\left[ p_2\beta y_0-2(E+E_\nu^0)
- E(1-\beta^2)\right] + Q_2\frac{p_2^2}{E}\right] \\
&  & \mbox{} - \frac{p_2l}{2}Q_4\frac{(E+E_\nu^0)^2}{E}-\frac{p_2l}{4M_1}E(1-\beta^2)B^- \\
&  & \mbox{} + \frac{p_2l}{2M_1}\left[ M_1\frac{h^+}{e}(E_\nu^0+\beta^2E+p_2\beta y_0)-E_\nu^0(1-\beta x_0)g_1
(g_2-f_2) - E(1-\beta^2)A^-\right],
\end{eqnarray*}
\begin{eqnarray*}
H_8^\prime & = & \frac{p_2l}{4}\left[ Q_1+Q_3-Q_2(2E+E_\nu^0)-Q_4(E-E_\nu^0)\right] + \frac{p_2l}{4M_1}(E-2E_\nu^0)B^-
\\
&  & \mbox{} + \frac{p_2l}{2M_1}\left[ E_\nu^0M_1\frac{2h^--h^+}{e}-E_\nu^0(2f_1f_2-g_1g_2-g_1f_2)+EA^-\right],
\end{eqnarray*}
\begin{equation*}
H_9^\prime = \frac{p_2\beta}{8}\left[ -Q_1-Q_3+(Q_2  +Q_4)(E+E_\nu^0)\right],
\end{equation*}
\begin{eqnarray*}
H_{10}^\prime & = & -\frac{p_2l^3}{4}(Q_2+5Q_4) + \frac{p_2l^3}{2M_1}(2B^-+C^-) \\
&  & \mbox{} + \frac{p_2l^3}{M_1}\left[-2M_1\frac{h^+}{e}-3M_1\frac{h^-}{e}+3f_1f_2+4g_1g_2-3g_1f_2 \right].
\end{eqnarray*}
It turns out that $H_{11}^\prime=0$ in this rearrangement. However, the indexing of Eq.~(5) is made easier by keeping
this $H_{11}^\prime$ explicitly.
\begin{equation*}
H_{12}^\prime = -\frac{p_2^2l^2}{2}Q_4+\frac{p_2^2l^2}{2M_1}B^++\frac{p_2^2l^2}{M_1}\left[ M_1\frac{h^+}{e} +A^+
-g_1(g_2-f_2)\right],
\end{equation*}
\begin{equation*}
H_{13}^\prime = \frac{p_2^2l^2}{2}Q_4-\frac{p_2^2l^2}{2M_1}C^--\frac{p_2^2l^2}{M_1}\left[ M_1\frac{h^+}{e}+f_2
(f_1+g_1)\right],
\end{equation*}
\begin{equation*}
H_{14}^\prime = -\frac{p_2l^2}{4}Q_2+\frac{p_2l^2}{4M_1}(B^-+2A^-),
\end{equation*}
\begin{equation*}
H_{15}^\prime = -\frac{p_2l}{8}(Q_2+Q_4)+\frac{p_2l}{4M_1}B^-+\frac{p_2l}{2M_1}\left[ -M_1\frac{h^-}{e}+f_2(f_1-g_1)
\right],
\end{equation*}
and
\begin{equation*}
H_{16}^\prime = \frac{p_2\beta}{4M_1}\left[ -M_1\frac{h^+}{e}+g_1(g_2  -f_2  )\right],
\end{equation*}
with
\begin{eqnarray*}
A^\pm & = & f_1f_2-g_1g_2 \pm 2g_1f_2, \\
B^\pm & = & (f_1\pm g_1)^2+2f_1(f_3-f_2), \\
C^- & = & f_1^2-g_1^2+2f_1(f_3-f_2), \\
h^\pm & = & -g_1^2(\kappa_1+\kappa_2) \pm f_1g_1(\kappa_2-\kappa_1).
\end{eqnarray*}
Here, $e$ is the electron charge, $\beta =l/E$, $y_0=\{(E_\nu^0)^2-l^2-p_2^2\}/(2lp_2)$, $x_0=-(p_2 y_0+l)/E_\nu^0$
and $\kappa_1$ and $\kappa_2$ are the anomalous magnetic moments of the decaying and emitted baryons, respectively.

The $\rho_i$ functions of Eq.~(\ref{CNAp}) are
\begin{equation*}
\rho_{N1} = \rho -\frac{l^2p_2^2}{2M_1}\theta_{13}
\end{equation*}
and
\begin{equation*}
\rho_{N2} = \rho + \frac{p_2l}{2M_1}\left[-(l^2+4p_2ly_0)\theta_4-4lE_\nu^0\theta_5+\beta p_2 \theta_{17}+2E_\nu^0
(1+y_0)\theta_{18}-3l^2\theta_{10}\right],
\end{equation*}
with
\begin{eqnarray*}
\rho & = & \frac{p_2l}{2M_1}(E_\nu^0-E)\theta_0+m^2(\theta_3-\theta_4)-l(E+2E_\nu^0)\theta_5+\left[ \beta p_2y_0+E
+E_\nu^0+\frac{m^2}{E}\right] \frac{\theta_7}{2} \\
&  & \mbox{} + (E_\nu^0-2E) \frac{\theta_8}{2}-3l^2\theta_{10}+\beta p_2(2E-E_\nu^0)\theta_{12}-l\theta_{14}
-\frac{\theta_{15}}{2}+\frac{\theta_{16}}{4E}.
\end{eqnarray*}

The $Q_i$ and $\theta_i$ functions are found in Refs.~\cite{tun91,tun93}.

\end{document}